\documentclass[prb,twocolumn,showpacs,10pt,showkeys,preprintnumbers,amsmath,amssymb]{revtex4-1}

\usepackage{graphicx}
\usepackage{dcolumn}
\usepackage{bm}
\usepackage{verbatim}
\usepackage[version=3]{mhchem}
\usepackage{subfigure}
\graphicspath{{images/}}

\begin{document}

\preprint{}

\title{Hillock formation of Pt thin films on Yttria stabilized Zirconia single crystals}

 \author{Henning Galinski}
 \email{henning.galinski@mat.ethz.ch}
 \author{Thomas Ryll}
 \author{Lukas Schlagenhauf}
 \author{Ludwig J. Gauckler}
 \affiliation{Nonmetallic Inorganic Materials, ETH Z\"urich, Z\"urich, Switzerland}
 
 \author{Patrick Stender}
 \author{Guido Schmitz}
 \affiliation{Institute for Materials Physics, WWU M\"unster, M\"unster, Germany}

\date{\today}

\begin{abstract}
The stability of a metal thin films on a dielectric substrate is conditioned by the magnitude of the interactive forces at the interface. In the case of a non-reactive interface and weak adhesion, the minimization of free surface energy gives rise to an instability of the thin film. In order to study these effects, Pt thin films with a thickness of 50 nm were deposited via ion-beam sputtering on yttria stabilized zirconia single crystals. All Pt films were subjected to heat treatments up to 973 K for 2 h. The morphological evolution of Pt thin films has been investigated by means of scanning electron microscopy (SEM), atomic force microscopy (AFM) and standard image analysis techniques. Three main observations have been made:
i) the deposition method has a direct impact on the morphological evolution of the film during annealing. 
Instead of hole formation, that is typically observed as response to a thermal treatment, anisotropic pyramidal shaped hillocks are formed on top of the film.
ii) It is shown by comparing the hillocks' aspect ratio with finite element method (FEM) simulations that the hillock formation can be assigned to a stress relaxation process inside the thin film.
iii) By measuring the equilibrium shapes and the shape fluctuations of the formed Pt hillocks the anisotropy of the step free energy and its stiffness have been derived in addition to the anisotropic kink energy of the hillock's edges.

\end{abstract}
\maketitle


\section{\label{sec:level1}Introduction}

Thin metal films on dielectric substrates are thermodynamically instable. Their stability has been subjected to research under several aspects:  thermodynamics and kinetics \cite{Galinski1,Srolovitz1,Genin1}, mass transport via surface diffusion \cite{Galinski1,BrandonBradshaw1,GontierMoya1}, impact of surface energy anisotropies \cite{Bussmann1,Dufay1,Stoecker1,Dornel1}, fingering instabilities \cite{Dufay1,Jiran1,Kan1}, Ostwald ripening of islands \cite{Kennefick1}, hole pattern \cite{Shaffir1} and hillock formation\cite{Frolov1,Hwang1,Gladkikh1,Sharma1,Nam1,Boettinger1}.Most of the fundamental theoretical work has been carried out by Srolovitz and Safran who developed a complete stability theory for thin films covering kinetics\cite{Srolovitz3} and energetics\cite{Srolovitz2}.\\
In the case of strained layers, the shape instability leads either to equilibrium-shaped hole or to equilibrium shaped hillock formation~\cite{Tersoff1,Tersoff2,Pimpinelli1} depending on the competing relaxation mechanism. While the formation of hillocks as a consequence of stress relaxation is observed for various thin film materials~\cite{Tersoff2,Vailionis1,Boettinger1,Frolov1,Iwamura1,Kim1,Lahiri1}, the formation of equilibrium shaped holes has been affirmed recently both experimentally and by kinetic Monte Carlo simulations~\cite{Bussmann1}. However, the nature and transition between these two competing instability mechanisms is far from being fully understood. This is mostly due to a lack of experimental data.\\ 
The main objective of this paper is to establish a relation between the observable macroscopical changes during hillock formation on a strained thin film and its underlying configurational forces, e.g., the kink energy. The investigation will focus on how the presence of an internal stress field in the thin film triggers the thin film instability caused by hillock formation. In a second step, it is addressed, how the equilibrium hillock shapes can be used to determine critical stability-related quantities like the step line stiffness $\tilde{\beta}$ and the kink energy $\epsilon$~\cite{Dufay1}.
Therefore standard finite element modeling (FEM) has been chosen alongside the analysis of the anisotropic hillock shape fluctuations, which is successfully applied to determine the step energies of 2D islands during thin film growth~\cite{Kodambaka1,Khare1,Ibach1}.\\
In the present study, Pt thin films on single crystalline yttria stabilized zirconia (\ce{ZrO2}) have been chosen as model metal/ceramic systems in order to study the hillock formation. In equivalence to Au/\ce{ZrO2} \cite{Renaud1}, the Pt thin film can be regarded as strained and its interface to \ce{ZrO2} as semicoherent~\cite{Beck1}, due to a lattice parameter misfit of $\epsilon_{m}=(a_{f}-a_{s})/a_{s}=0.31$.\\
This paper is structured as follows: Section~\ref{sec:level2} derives the basic theoretical models necessary to describe the formation of hillocks. Section~\ref{sec:level3} deals with the experimental framework. In Section~\ref{sec:level4}, detailed results are presented and discussed. The final section~\ref{sec:level5} encompasses a summary of the findings and conclusions.

\section {\label{sec:level2}Theoretical Background}

\begin{figure}[t]
  \begin{center}
    \subfigure[~dense]{\label{fig:scheme-a}\includegraphics[scale=0.22]{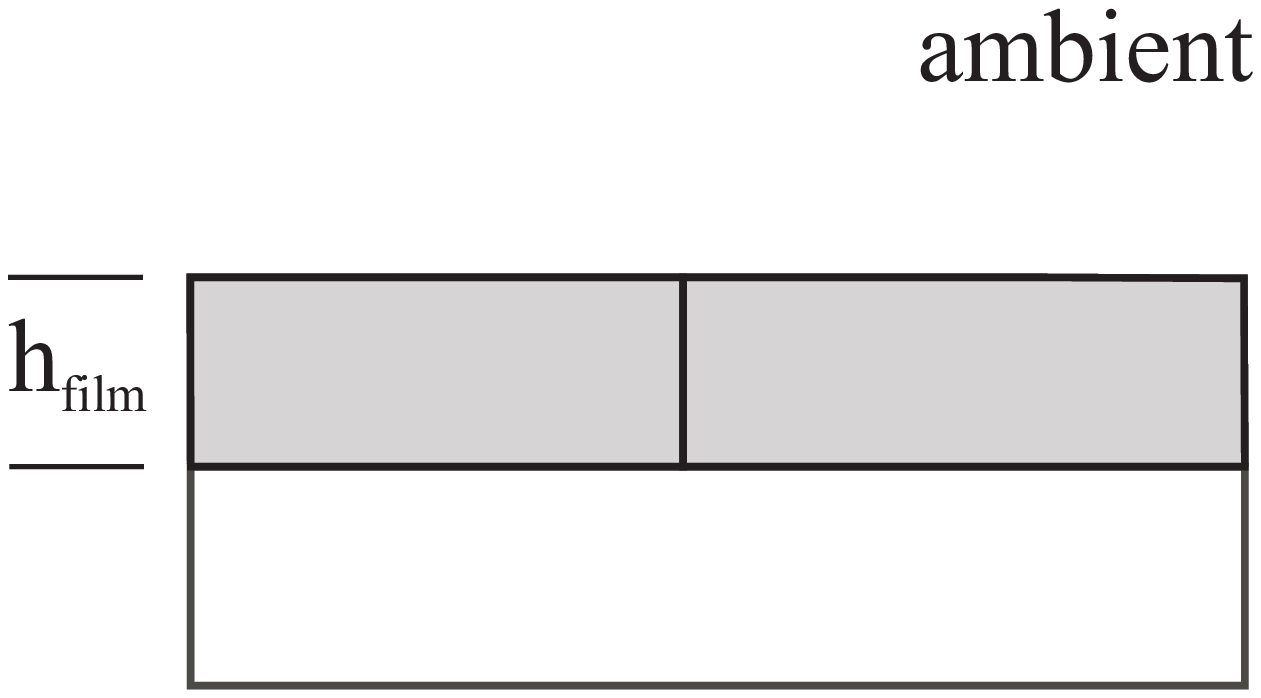}}
    \subfigure[~hillock]{\label{fig:scheme-b}\includegraphics[scale=0.22]{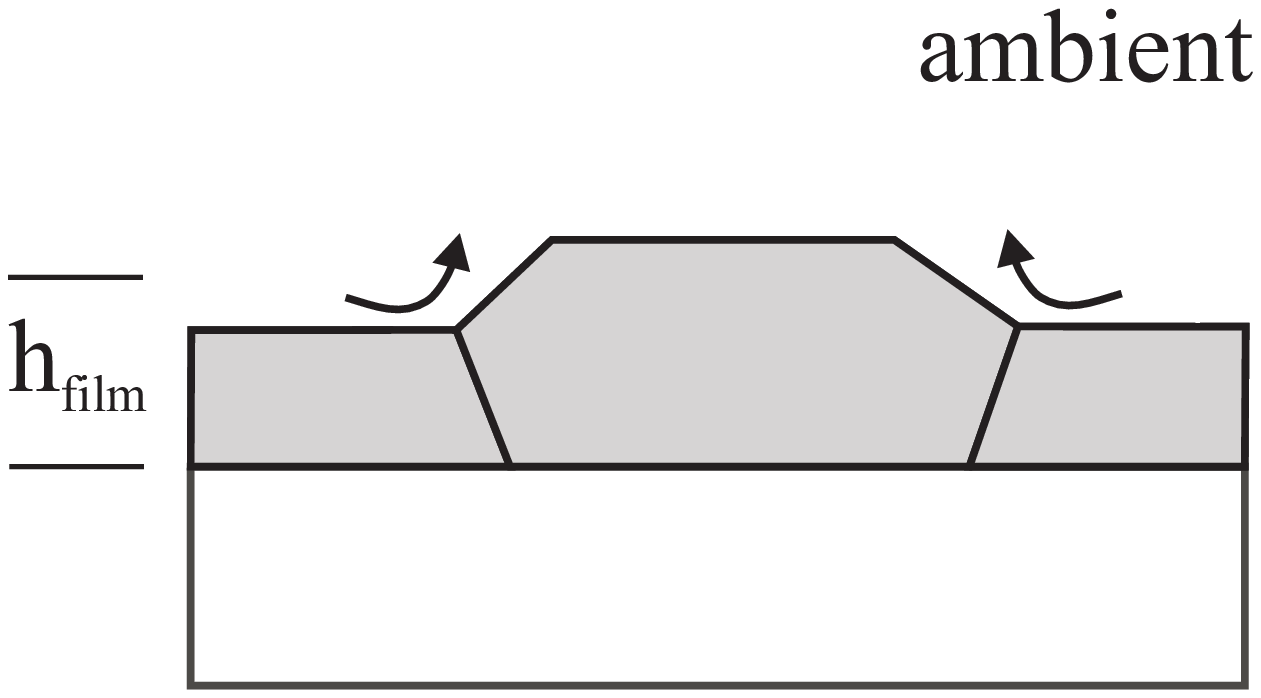}}
    \subfigure[~hillock+holes]{\label{fig:scheme-c}\includegraphics[scale=0.22]{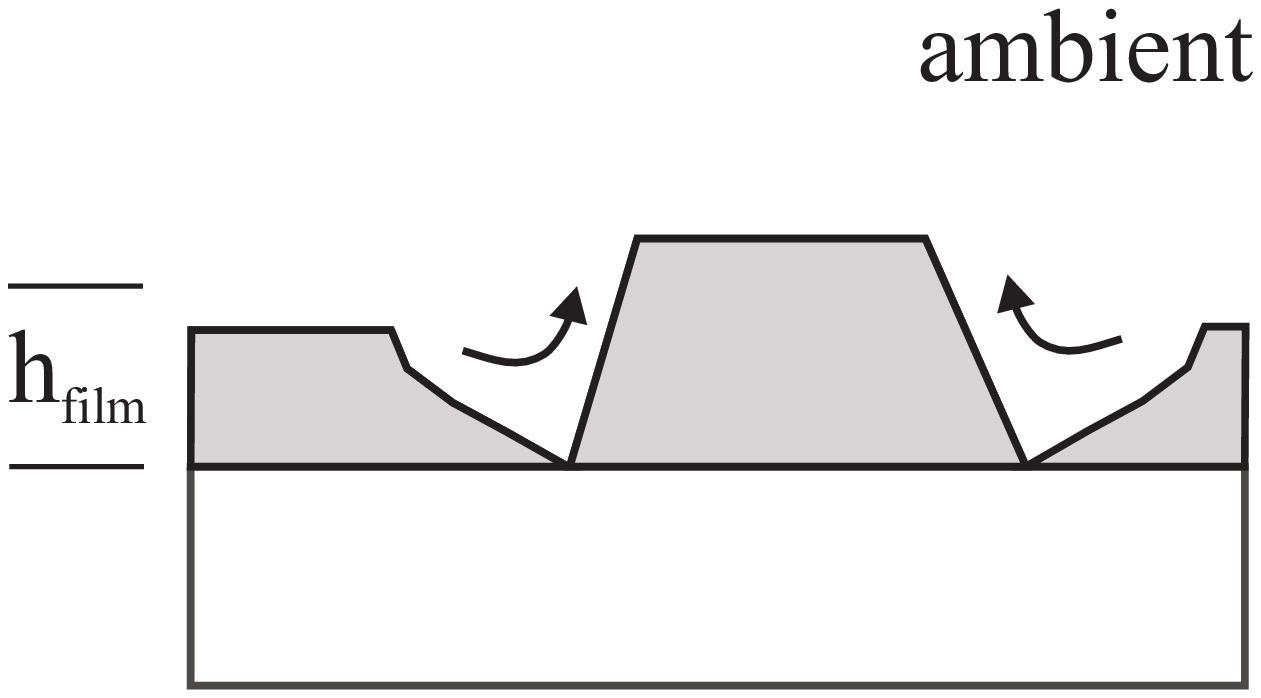}}
  \end{center}
  \caption{Illustration of the experimentally observed regimes of hillock formation (a) during thermal treatment of a dense flat film (b) hillocks form on the film surface with a regular hexagonal shape (c) due to further annealing holes and secondary hillocks form in the vicinty of the primary hillock}
  \label{fig:Scheme}
\end{figure}

\subsection{\label{sec:level21}Basics}

During deposition of metal thin films via sputtering, the kinetic energy $E_{\text{kin}}$ of the deposited atoms generally exceeds their thermal energy $E_{\text{therm}}$. For the deposited film, this results in a metastable configuration that tends to equilibrate, once subjected to temperature by annealing, Joule's heating or radiation. The morphological stability of a metal thin film on a dielectric material is thus conditioned by the aspect ratio, the interaction across the interface~\cite{Hosson1} and the tendency of the thin film to reduce its free energy, e.g., due to stress relaxation. 
The way in which the reduction of the free energy manifests in the evolution of the film morphology depends strongly on the competing relaxation mechanisms. While in typical thin film agglomeration scenarios, defect related local perturbations cause a film rupture and a decrease of surface area~\cite{Galinski1}, the formation of hillocks on the contrary is the direct response to a delocalized strain field in the thin film. The origin of this strain field is attributed to a lattice mismatch $\epsilon_m$ or a growth stress induced by the deposition technique~\cite{Debelle1}. It is noteworthy, that in contrast to thin film agglomeration the formation of the hillock is usually accompanied by the increase of the surface area as shown schematically in Fig.~\ref{fig:Scheme}. The change of the total free energy $\Delta F$ of a uniformly strained film with volume $V$ to a (partially) relaxed film with hillock on top can be expressed in terms of the change in strain energy density $\Delta W$ and surface energy $\Delta\Phi$,
\begin{equation}
\Delta F = \Delta W+ \Delta\Phi=(W-W_0)+(\Phi-\Phi_0).
\label{eq:HIL1}
\end{equation}
Whereby $\Delta\Phi >0$, if only hillocks are formed. Hence, the reduction of the free energy can only be caused by a decrease in elastic energy which scales with the initial elastic energy  
\begin{equation}
W_0 = V\!M\!\epsilon^{2}_{m},
\label{eq:HIL2}
\end{equation}
where $M=\frac{(1-\nu)}{2\pi\mu}$ with the Poisson ratio $\nu$ and the shear modulus $\mu$. The relaxed strain energy $W$ throughout the film volume $V$ is considered to be equal to the induced change in the elastic strain field $\epsilon_{ij}$ by the formation of hillocks. Thus   
\begin{equation}
W = \int_V \frac{1}{2}c_{ijkl}\epsilon_{ij}\epsilon_{kl}\mathrm{d}V,
\label{eq:HIL3}
\end{equation}
where $c_{ijkl}$ is the stiffness tensor of the material. In general, Eq.~\ref{eq:HIL3} has to be solved numerically.\\ 
\begin{figure}[h!]
	\centering
		\includegraphics[scale=1.20]{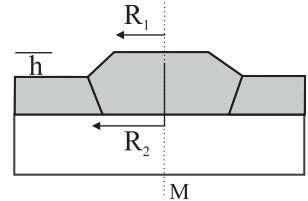}
	\caption{Schematic drawing of an elastically strained film including a hillock in the shape of a frustum, rotationally symmetric around the axis $M$, with the top radius $R_1$, the bottom radius $R_2$ and the height $h$}
	\label{fig:hillock formation_fem_1}
\end{figure}
In order to facilitate the calculation, the three dimensional hillock can be approximated by a two dimensional frustum with rotational symmetry, as shown in Figure~\ref{fig:hillock formation_fem_1}. By doing so, the calculation of the change in surface energy 
\begin{equation}
\Delta\Phi= \gamma_f\Omega-\gamma_f\Omega_0
\label{eq:HIL5}
\end{equation}
is simplified as it is characterized only by the uniform surface energy $\gamma_f$ and the surface area $\Omega$ given by the three geometrical parameters $R_1$,~$R_2$ and $h$. The surface after a hillock has formed reads,
\begin{equation}
\Omega=\pi(R_1+R_2)\cdot\sqrt{R^2_1-2R_1R_2+R^2_2+\acute{h^2}}+\pi R^2_1+(\Omega_0-\pi R^2_2).
\label{eq:HIL5}
\end{equation}
$W$ obviously depends on the film morphology and therefore on $R_1$,$R_2$ and $h$. $\Omega_0$ is the surface area of the initially flat film.

\subsection{\label{sec:level23}Hillock Shape Analysis}

Hillock formation as well as thin film agglomeration require the motion and creation of atomic steps. Thereby the general anisotropic line stiffness $\tilde\beta$ of the atomic steps, serves as the key parameter to investigate and understand the morphological evolution of a wide class of thin film instabilities~\cite{PierreLouis1}. 
In analogy to islands growth~\cite{Kodambaka1,Michely1}, the threefold symmetric hillocks can be treated as two dimensional shapes and their contour line can be expressed in terms of a Fourier-series with the general form
\begin{equation}
R_{\text{fit}}(\theta)=R_0+\sum^{3}_{i=1}a_i\cdot\sin(n_i(\theta-\theta_i)).
\label{eq:ECS2}
\end{equation}
Whereby $R_0$,~$n_i$,~$a_i$,~$\theta_i$ serve as fitting parameters. The step free energy per unit length $\beta$, or step line tension, is related to the hillock shape $R_{\text{fit}}$ by the 2D Wulff construction~\cite{Pimpinelli1}, which has been proven by Burton \textit{et al.}\cite{Burton1} to be
\begin{equation}
\beta(\phi)=\lambda \frac{R^2_{\text{fit}}}{\sqrt{R^2_{\text{fit}}+\mathrm{d}_{\theta}R^2_{\text{fit}}}}.
\label{eq:ECS1}
\end{equation}
Thereby $\phi=\theta-\arctan\left({\mathrm{d}_{\theta}R_{\text{fit}}/R_{\text{fit}}}\right)$ denotes the normal to the equilibrium shape for each $R_{\text{fit}}(\theta)$. It is noteworthy, that Eq.~\ref{eq:ECS1} establishes the proportionality between the measurable 2D shape $R_{\text{fit}}(\theta)$ and its free energy $\beta$. \\
For small shape fluctuations, the elongation of the step contour line has to be taken into account and the step line tension is replaced by the step line stiffness $\tilde{\beta}(\phi)=\beta(\phi)+\mathrm{d}_{\phi\phi}\beta$. Similar to Eq.~\ref{eq:ECS1}, the step line stiffness $\tilde\beta$ is orientation-depended and related to the hillock's curvature $\kappa$ by 
\begin{equation}
\tilde{\beta}(\theta)=\lambda\cdot\left[\frac{(R^2_{\text{fit}}+\dot{R^2_{\text{fit}}})^{\frac{3}{2}}}{R^2_{\text{fit}}+2\dot{R^2_{\text{fit}}}-\ddot{R_{\text{fit}}}}\right]=\frac{\lambda}{\kappa(\theta)}.
\label{eq:ECS3}
\end{equation}
Once the equilibrium shape of a hillock is reached, the curvature $\kappa$ of steps along the densely packed directions is zero and $\tilde{\beta}=\beta$. Conversely, all shape fluctuations cause a curvature $\kappa$ of the step. Using the Terrace Step Kink (TSK) model~\cite{Ibach1}, this curvature is related to the thermally activated formation of kinks along the step with the length of $n$ atomic units $a_{\|}$. In the case of an unrestricted TSK-model, the step line stiffness $\tilde{\beta}$ is given by    
\begin{equation}
\tilde{\beta}=(2\:a_{\|}\:k_b\:T/a^{2}_{\bot})\sinh^2{\frac{\epsilon_{k}}{2\:k_b\:T}},
\label{eq:ECS5}
\end{equation}
where $\epsilon_{k}$ is the required kink formation energy and $a_{\|}=0.277$~nm and $a^{2}_{\bot}=0.240$~nm are the unit lattice spacing parallel and orthogonal to the step edge, respectively.\\
The only variable in the upper equations, which is not yet determined, is $\lambda$. Kodambaka \textit{et al.}~\cite{Kodambaka2} established a generalized formulation for anisotropic 2D crystal shapes, which directly relates $\lambda$ and the experimentally accessible hillock shape fluctuation function $g$. The hillock shape fluctuations $g$ are determined by calculating the deviation of the experimentally measured hillock contour from the fitted equilibrium shape given by Eq.~\ref{eq:ECS2} and is defined as follows,      
\begin{equation}
g(\theta)\equiv(r_{\text{exp}}-R_{\text{fit}})/R_{\text{fit}}.
\label{eq:ECS6}
\end{equation}
In order to relate $g$ to $\lambda$, two fluctuation sensitive functions 
\begin{equation}
\chi(\theta)=\frac{R^{2}_{\text{fit}}\mathrm{d}_{\theta}g}{\sqrt{R^{2}_{\text{fit}}+2\mathrm{d}_{\theta}R^{2}_{\text{fit}}-R_{\text{fit}}\mathrm{d}_{\theta\theta}R_{\text{fit}}}}
\label{eq:ECS8}
\end{equation} 
and
\begin{equation}
\rho(\theta)=g\cdot R_{\text{fit}}
\label{eq:ECS9}
\end{equation}
are defined. The analytical functions of $\chi$ and $\rho$ are expressed in terms of a Fourier series with Fourier coefficients $\chi_n$ and $\rho_n$. By applying the equipartion theorem, which states that in equilibrium every DOF has the same energy $\left\langle E \right\rangle=\frac{1}{2}k_b T$, it can be shown that    
\begin{equation}
\lambda(g)=\frac{N_{\text{max}} k_b T}{2\pi\sum_{n}{\left\langle \left|\chi_n\right|^2-\left|\rho_n\right|^2\right\rangle}},
\label{eq:ECS10}
\end{equation}
where $N_{\text{max}}=0.5\cdot N_{\text{atoms},\partial\Omega}$ is defined as one half of all atoms along the contour line $\partial\Omega$ of the hillock. A detailed reproduction is given in reference~\cite{Kodambaka2}.

\section {\label{sec:level3}Experimental}
\subsection{\label{sec:level32}Sample Preparation and Characterization}
The Pt/\ce{ZrO2}(single-crystalline) system is immiscible and characterized by a lattice mismatch of $\epsilon_{m}=(a_{f}-a_{s})/a_{s}=0.31$. A chemical inert interface is formed for all temperatures below $1273$~K~\cite{Lu1}. Pt layers of 50~nm in thickness were deposited at room temperature by ion beam sputtering
($p_{\text{base}}=1\cdot10^{-7}$~mbar) upon the single-crystalline substrates that were pre-cleaned using isopropanol. Before deposition, the single crystalline substrates were cleaned in the Ar-ion beam of the sputtering chamber for 5~s. The substrates coated with the thin platinum film were annealed in a muffel furnace for 2~h at 923~K and 973~K. The annealing temperatures are well below the melting temperature of platinum $T_{M}=2042$~K, hence volume diffusion of Pt in Pt is prohibited.
The morphology of the samples was studied via high resolution AFM, using a Mobile S (Nanosurf), and SEM (Zeiss Leo 1530).  For all acquired hillocks, the aspect ratio $a$ and the equilibrium shape $r(\theta)$ have been determined.

\subsection{\label{sec:level32}Finite Element Modelling}

In order to confirm the role of incompatible strains due to a lattice mismatch as the primary reason for hillock formation, a thermoelastic FEM model using COMSOL Multiphysics has been developed. The film is assumed to be subjected to an extensional mismatch strain $\epsilon_m$. The origin of this strain is the large lattice mismatch $f$ between film and substrate which can be treated as a thermoelastic deformation. A nonuniform elastic strain $\epsilon_{i,j}$ is created by assigning a thermal expansion strain equal to $-\epsilon_m$ in the thin film. A frustum is chosen to represent the hillock shape in first approximation, see Fig.~\ref{fig:hillock formation_fem_1}. The frustum is rotationally symmetric around the axis $M$ and features therefore three-dimensional effects. In analogy to the AFM-analysis, the frustum's bottom radius $R_2$ and the hillock height $h$ define the hillock's aspect ratio $a=h/R_2$.
The film is treated to be isotropic, hence the initial strain energy is $M_f\:\epsilon_m\:V$ where $M_f$ is the biaxial modulus of the film and $V$ the total film volume. The total free energy change $\Delta F$ of the system during hillock formation is simulated starting from a flat film with increasing hillock height $h$ and coevally decreasing bottom radius $R_2$. The material-specific constants used in the FEM model are listed in Tab.~\ref{tab:FEM-Paramters}.

\begin{table}[h!]
	\caption{\label{tab:FEM-Paramters} FEM model parameters for the elastic constants $E$, $\nu$ and the surface energies $\gamma_{\text{Pt}}$, $\gamma_{\text{YSZ}}$ and $\gamma_{\text{Pt/YSZ}}$}
\begin{ruledtabular}
\begin{tabular}{ccc}
Parameter & Value & Lit. \\
\hline
$E$&$1.63\cdot10^{11}\text{~Pa}$&\cite{Davis1}\\
$\nu$&$0.390$&\cite{Davis1}\\
\hline
$\gamma_{\text{Pt,111}}$&$1.656$&\cite{Zhang1}\\
$\gamma_{\text{YSZ}}$&$1.927$&\cite{Tsoga1}\\
$\gamma_{\text{Pt/YSZ}}$&$1.2$&\cite{Galinski1}\\
\end{tabular}
\end{ruledtabular}
\end{table}

\section {\label{sec:level4}Results and Discussion}

\subsection{\label{sec:level41}Basics}

\begin{figure}[t!]
  \begin{center}
    \subfigure[~923~K]{\label{fig:hillock_sem-a}\includegraphics[scale=1.4]{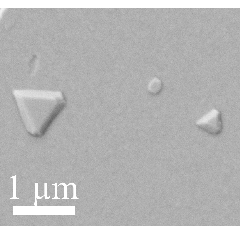}}
    \hspace{0.5cm}
    \subfigure[~973~K]{\label{fig:hillock_sem-b}\includegraphics[scale=1.4]{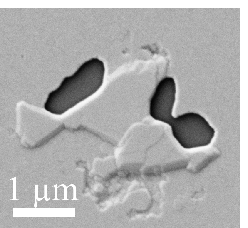}}
  \end{center}
  \caption{SEM images of different stages of hillock formation on Pt(111) with corresponding annealing temperature. (a) Hillock formation on top of the dense film, (b) Additional hillock growth and holes forming in the vicinity of the hillock.}
  \label{fig:hill}
\end{figure}

\begin{figure}[h!]
  \begin{center}
    \subfigure[]{\label{fig:hillock-a}\includegraphics[scale=1.0]{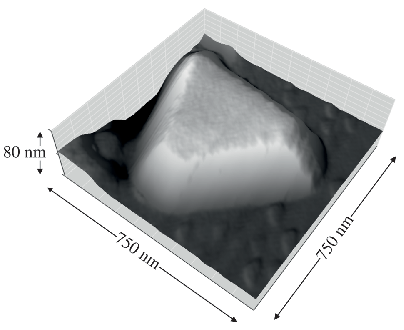}}
    \subfigure[]{\label{fig:hillock-b}\includegraphics[scale=0.22]{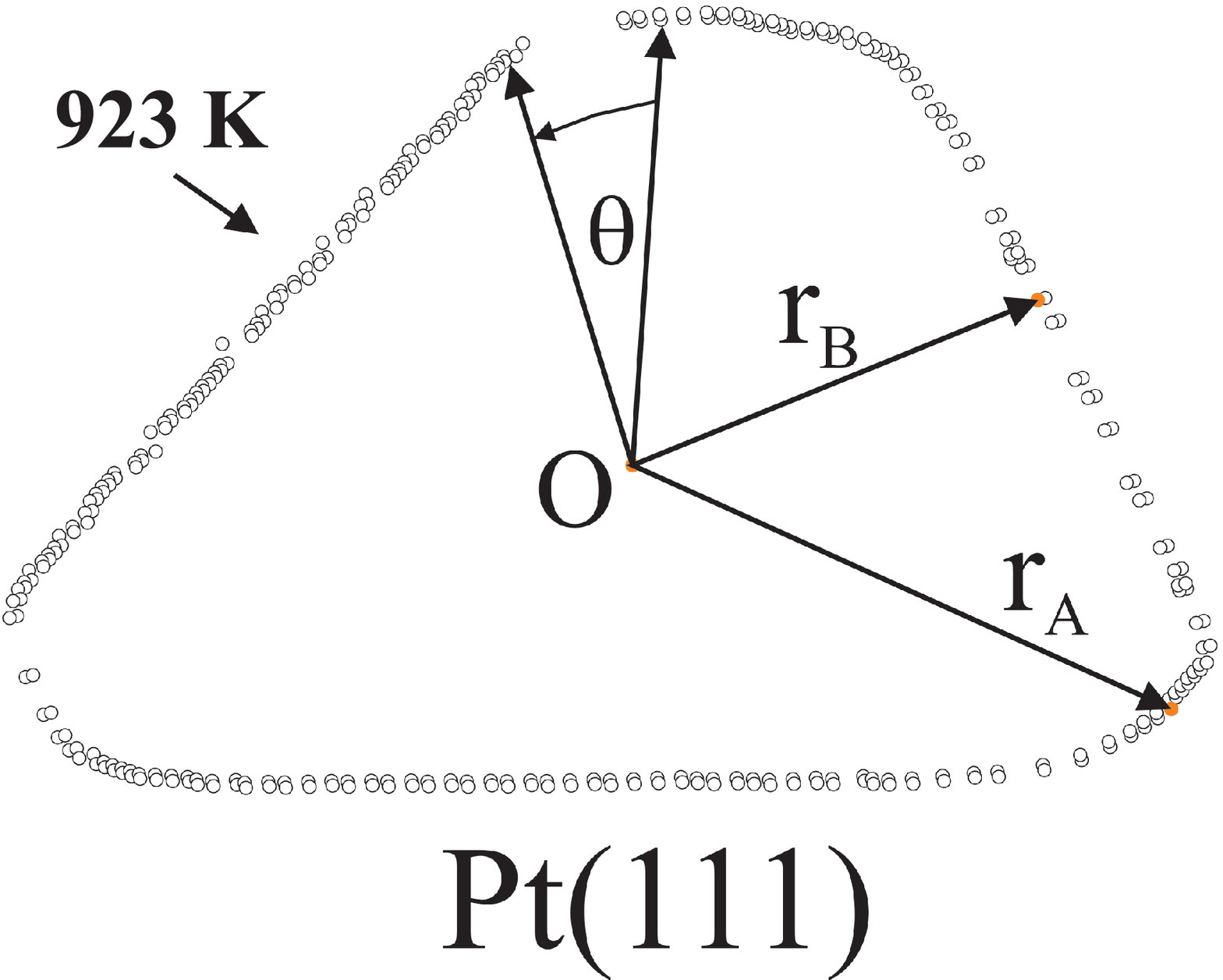}}
  \end{center}
  \caption{(a) 3D AFM image ($750\cdot 750$nm$^2$) of a hillock on Pt(111) annealed at 923~K. (b) Hillock perimeter of the formed hillock obtained via standard image analysis techniques, the differing radii $r_a$ and $r_b$ indicate a anisotropy in the line tensions for A- and B-steps.} 
  \label{fig:edge}
\end{figure}

For all isochronically annealed $(t=2~h)$ Pt thin films, the stages of hillock formation have been analyzed. In Fig.~\ref{fig:hill}, the two different stages of hillock formation on the Pt thin film with an initial thickness $h=50$~nm are shown representatively. For sufficiently low annealing temperatures $T_a$, the film relaxation is solely driven by the formation of hillocks while all other regions on the film are smooth, see Fig.~\ref{fig:hillock_sem-a}. Energetically speaking the gain in free energy is dominated by the strain relaxation, hence $W<\Phi$. The observed hillock shapes in Fig.~\ref{fig:hillock_sem-a} are highly anisotropic and characterized by a truncated triangle with a threefold symmetry. This symmetry is defined by the different edge lengths $\left(A,B\right)$ of the hillock which entail the energetical differences of the series of steps forming these edges. The  base of the hillock is enlarged in respect to its height $h$, which reflects the expected kinetics for hillock growth described by Tersoff and Tromp~\cite{Tersoff1}. In contrast to recent molecular dynamics simulations of hillock growth~\cite{Frolov1}, the hillock nucleation in this work does not demand a seed grain positioned on top of the film.\\ 
By increasing $T_a$, the morphological evolution of the thin film undergoes a transition indicated by the coexistience of two different relaxation modes: the additional hillock formation and film rupture in the vicinity of the hillock [Fig.~\ref{fig:hillock_sem-b}].  Thereby the film rupture along the long edges ($B$-steps) of the hillock corresponds to the onset of surface roughening in regions of local stress enhancement, as shown in Fig.~\ref{fig:hillock_sem-b}. However, the additional aggregation of hillocks preferentially occurs at the short edges of the hillock, which coincides with the low energy steps ($A$-steps). While the hillocks base is significantly enlarged, the mean hillock height is not severely affected by the 
heat treatment. These observations are in good agreement with the predictions for shape changes induced by strain relaxation made by Tersoff~\cite{Tersoff1}.\\
The hillock shapes and their aspect ratio $a$ have been acquired by high resolution atomic force microscopy, with a lateral resolution $< 0.2$~nm. In total $35$ hillocks were analyzed. In Fig.~\ref{fig:hillock-a} a typical $80$~nm high hillock is shown. The hillock has a characteristic threefold symmetry, the edges of the hillock are steep. The base radius $R_2$ and top radius $R_1$ of the hillock differ slightly. For all hillocks, formed after annealing at $T_a=923$~K, the contour line of the base has been measured and the center $O$ has been calculated. Furthermore the radii $r_A$ and $r_B$ corresponding to the $A$ and $B$ steps have been determined, see Fig.\ref{fig:hillock-b}. Due to the different orientation of the steps $A_{(100)}$ and $B_{(111)}$, the step formation energies are generally different. The anisotropy manifests in the ratio of step formation energies, which results from the Wulff-relation $\beta_{A}/\beta_{B}=r_{A}/r_{B}$. The Wulff relation follows from Eq.~\ref{eq:ECS1} provided that $\mathrm{d}_{\theta}R=0$, and thus applies only to the extrema of the hillock radius $r$. The mean ratio of the step free energy resulting from the measured $A$ and $B$ radii is $\beta_{A}/\beta_{B}=1.56(11)$ and given in Tab.~\ref{tab:AFM-Paramters1}. 
\begin{figure*}[t!]
  \begin{center}
    \subfigure[~$a=0.19$]{\label{fig:edge-b}\includegraphics[scale=1.3]{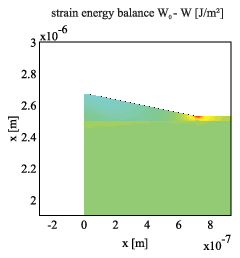}} 
    \subfigure[~$a=0.39$]{\label{fig:edge-c}\includegraphics[scale=1.3]{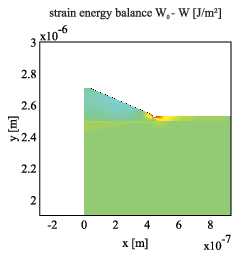}}
    \subfigure[~$a=0.70$]{\label{fig:edge-d}\includegraphics[scale=1.3]{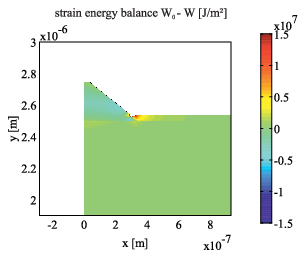}}
  \end{center}
  \caption{(a-c) Strain energy density balance calculated using a thermoelastic FEM-model of a $50$ nm Pt thin film on \ce{ZrO2} for rotationally symmetric hillocks with increasing aspect ratio $a$.}
  \label{fig:fem}
\end{figure*} 
The found anisotropy and energetically preferred formation of $B$-step is in accordance with both growth experiments on pit formation on Pt(111) single crystals~\cite{Michely1} and \textit{ab initio} calculations on Pt(111) step energies~\cite{Boisvert1}, see also Tab.~\ref{tab:AFM-Paramters1}.
\begin{table}[h!]
	\caption{\label{tab:AFM-Paramters1} Comparison of the determined length ratio $r_A/r_B$ of the hillock, and the step energy ratio $\beta_A/\beta_B$ with data from literature}
\begin{ruledtabular}
\begin{tabular}{ccc}
Parameter & Value & Lit. \\
\hline
$\beta_A/\beta_B$&$1.56(11)$&this work\\
$\beta_A/\beta_B$&$1.15(2)$&exp.\cite{Michely1}\\
$\beta_A/\beta_B$&$1.13$&theor.\cite{Boisvert1}\\
\end{tabular}
\end{ruledtabular}
\end{table}
However, the difference in the measured step energy ratio in this work and the ones of Michely \textit{et al.}~\cite{Michely1} and Boisvert~\cite{Boisvert1} is significant and can be related to an unequal mechanism that causes the relaxation process. While in the present work the relaxation is controlled by a stress field generated from the lattice mismatch between film and substrate, the formation of pits with anisotropic shape found in literature~\cite{Michely1}~\cite{Boisvert1} originates from a 2D to 3D growth transition, which is facilitated by a large tensile stress of the Pt(111)-surface.\\
This detailed impact on hillock formation of a mismatch induced stress field is addressed in the next section.  

\subsection{\label{sec:level42}Finite Element Modelling}

In consequence of the previous findings we assume that growth of hillocks is caused by a relaxation process whereby the minimization of the free energy $\Delta F$ is dominated by strain relaxation. Using the concepts derived in Sec.~\ref{sec:level21}, the experimentally revealed aspect ratio $a=h/R_1$ obtained by AFM are compared using FEM modeling with the predicted optimal trade-off between hillock shape and newly formed surface area $\Omega$. The total film volume is conserved, thus the measured effect is solely defined by a redistribution of mass. In Fig.~\ref{fig:fem}, the strain energy balance $W_0-W$  for three different aspect ratios $a$ is shown. In comparison to its initial value $W_0$, the strain energy density $W$ is decreased significantly (blue regions) by the formed hillock for all $a$ in Figure~\ref{fig:fem}. Furthermore with increasing $a$ an increasing strain energy density in the vicinity of the hillock base has been observed. As shown in Fig.~\ref{fig:edge}, these regions of high strain energy density coincide with regions of film rupture after further thermal treatment. This is due to the facilitated formation of  
\begin{figure}[b!]
  \begin{center}
    \subfigure[]{\label{fig:fem-a}\includegraphics[scale=0.22]{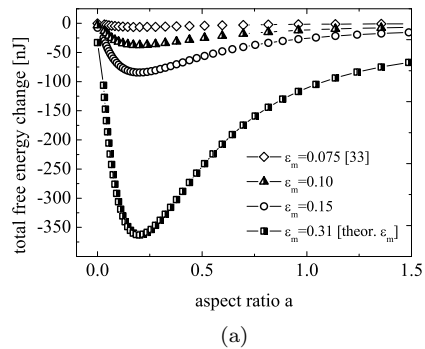}}\\
    \subfigure[]{\label{fig:fem-b}\includegraphics[scale=0.22]{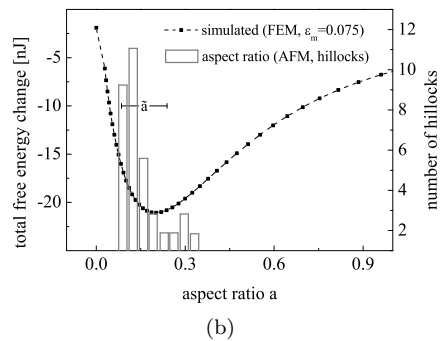}} 
  \end{center}
  \caption{(a) Total free energy change as function of the aspect ratio $a$ as determined from the FEM-model. (b) Evolution of the total free energy as function of the aspect ratio $a$ (FEM) including the experimentally determined aspect ratio distribution of hillocks (AFM).}
  \label{fig:hillock_fem}
\end{figure}
\begin{figure*}[t!]
  \begin{center}
    \subfigure[]{\label{fig:hf-a}\includegraphics[scale=1.75]{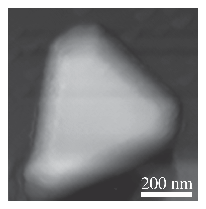}}
    \subfigure[]{\label{fig:hf-b}\includegraphics[scale=0.23]{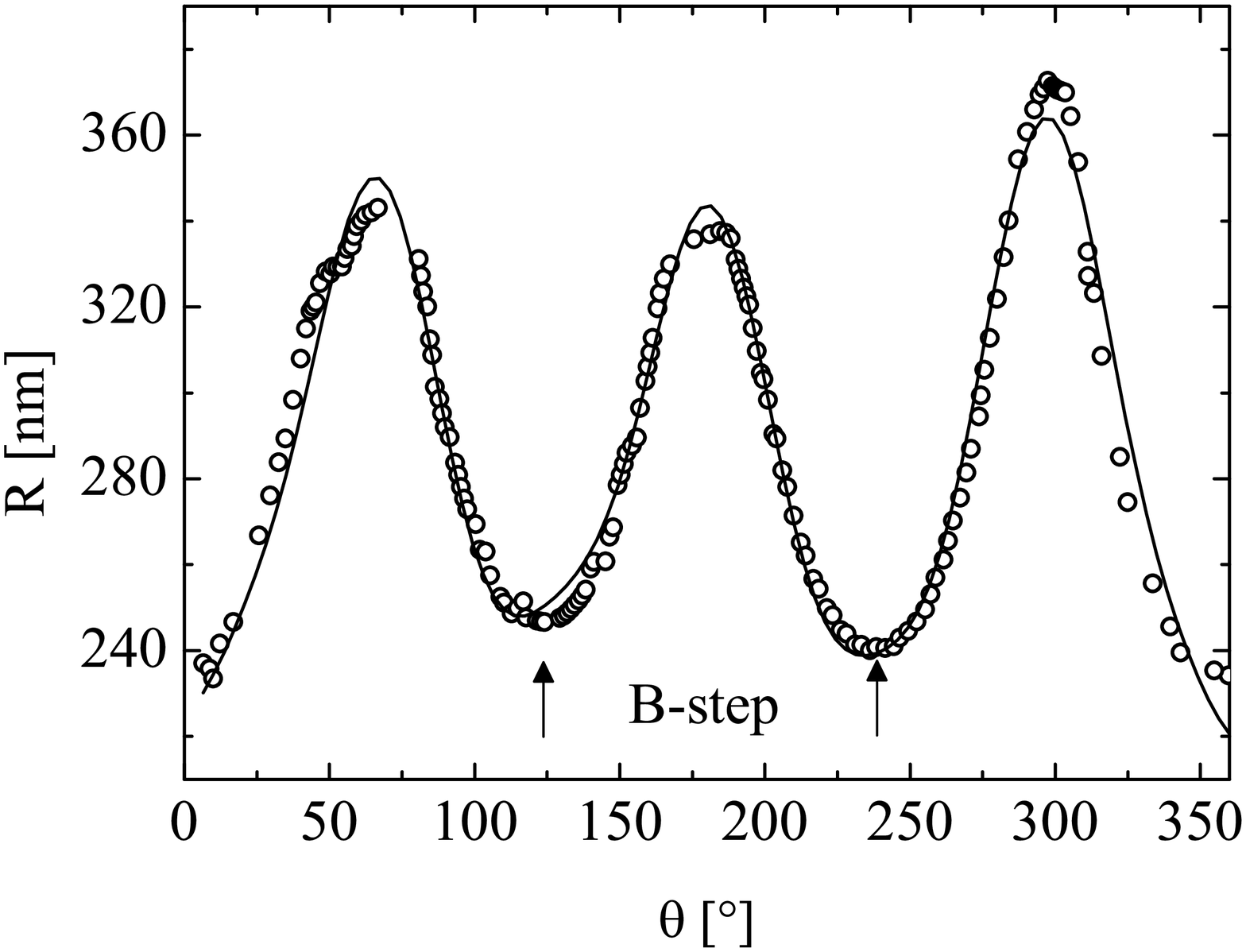}}
  	\subfigure[]{\label{fig:hf-c}\includegraphics[scale=0.96]{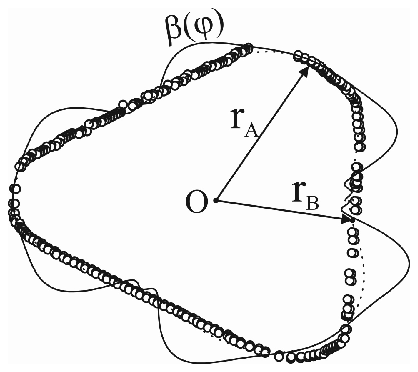}}   
  \end{center}
  \caption{(a) AFM image of a hexagonal shaped hillock on Pt(111) film annealed at 923~K (b) Measured equilibrium hillock shape plotted in polar coordinates radius $R$ vs. angle $\theta$ (open circles) fitted with Eq.~\ref{eq:ECS2} (solid line) (c) Polar plots of the fitted equilibrium shape $R\left(\theta\right)$ and the step free energy $\beta(\phi)$ with $\lambda=1$}
  \label{fig:hillock_fem}
\end{figure*}
dislocations and mass transport caused by the enhanced stress in the film~\cite{Tersoff2}.\\
The relaxation process has been modeled by FEM on the grounds of Eqs.~\ref{eq:HIL2}-\ref{eq:HIL3} for aspect ratios $a$ ranging from $0.1$~to~$2$ for four different lattice mismatches $\epsilon_m$ ranging from the theoretical lattice mismatch $\epsilon_m=0.31$ to a mismatch $\epsilon_m=0.075$ resulting from HRTEM analysis~\cite{Beck1}. It has to be noted, that the implemented strain in the film does not necessarily result from a lattice mismatch $\epsilon_m$ but can also be due to an intrinsic growth strain.
The total free energy change $\Delta F$ as function of $a$ and $\epsilon_m$ is shown in Fig.~\ref{fig:fem-a}. $\Delta F(a)$ resembles a Lennard-Jones potential, whose potential-well depth scales with $\epsilon_m$. However, the position of the minimal free energy stays unaffected from changes in $\epsilon_m$ at $a=0.20$, hence the found relation can be regarded as general scaling law under the condition that $\left|\Delta W\right|>\left|\Delta\Phi\right|$. \\
In order to validate the functional relationship between $\Delta F$ and $a$, the experimentally quantified distribution of hillock aspect ratios $a$ has been compared in Fig.~\ref{fig:fem-b} to the minimal free energy configuration predicted by the FEM modeling. Despite the geometrical simplification in the FEM model, a good agreement between the measured mean hillock aspect ratio $\tilde{a}=0.19(2)$ and the predicted minimal free energy configuration at $a=0.20$ of the thin film has been found. Thus, it can be concluded that the found shape change of the thin film is dominated by a strain relaxation process, whereby the formed hillocks correspond to the minimal energy configuration of the film.  
\subsection{\label{sec:level43}Hillock Shape Analysis}

In order to provide a deeper insight into the fundamentals of hillock formation, the shape fluctuation method~\cite{Ikonomov2,Kodambaka1,Schloesser1,Steimer1} is used to determine the kink energy $\epsilon$ and the corresponding line stiffness $\tilde\beta$. The anisotropic kink energy $\epsilon$ is a physical key property that impacts the physics and decay kinetics of two-dimensional Ostwald ripening~\cite{Kodambaka1}, island growth~\cite{Michely1,Kodambaka2}, vacancy island formation~\cite{Ikonomov2}, fingering instabilities~\cite{Dufay1}, thin film agglomeration~\cite{Bussmann1} and hillock formation.\\  
Due to the anisotropy of the found equilibrium hillock shapes, $\epsilon$ and $\tilde\beta$ depend on the step orientation $A,B$. By using the unrestricted TSK model (Eq.~\ref{eq:ECS5}), the kink energy can be determined from the line stiffness $\tilde\beta$, which is directly related to the chemical potential of the hillock~\cite{Ibach1}.\\
In the following discussion, the presented results apply exclusively to hillocks formed at $T=923$~K. At this temperature a hillock is not restricted or influenced by another hillock or hole in its vicinity. The shape fluctuations of hillocks with radii ranging from 150 to 550 nm have been analyzed. For each hillock, the equilibrium shape $R$ and $g$ were determined from the AFM data. A typical data set is shown in Fig.~\ref{fig:hillock_fem}. It is composed of the initially measured hillock by AFM (Fig.~\ref{fig:hf-a}), the educed polar-plot of the hillock perimeter fitted by Eq.~\ref{eq:ECS2} (Fig.~\ref{fig:hf-b}) and a combined mapping of the calculated step free energy (Eq.~\ref{eq:ECS1}) with $\lambda=1$ and the hillock shape in Cartesian coordinates (Fig.~\ref{fig:hf-c}). Due to entropy induced shape fluctuations~\cite{Khare1}, there is a measurable difference between the fit-function and experimental data in Fig.~\ref{fig:hf-b}. As long as the shape fluctuations $g$ are small, typically $<10\%$, the fluctuations of the step free energy $\delta\beta$ are small and hence can be approximated by the step line stiffness $\tilde{\beta}$ (see Eq.~\ref{eq:ECS5}). This condition is complied with all hillocks, as the chosen fit-function (Eq.~\ref{eq:ECS2}) agrees well with the experimental hillock shapes measured by AFM ($R^2>0.92$).\\
The shape fluctuation function $g(\theta)$ has been determined for each hillock from the AFM data. Subsequently $g(\theta)$ and $R_{\text{fit}}$ have been inserted into Eqs.~\ref{eq:ECS8} and~\ref{eq:ECS9} to calculate $\chi(\theta)$ and $\rho(\theta)$. The two functions have been expanded into Fourier series. The resulting Fourier coefficients $\chi_n$ and $\rho_n$ are used to obtain $\lambda$ for each hillock.
\begin{table}[h!]
	\caption{\label{tab:AFM-Paramters2} Comparison of the experimentally determined stiffness $a_{\|}\tilde{\beta_A}$ and $a_{\|}\tilde{\beta_B}$ and the kink energies $\epsilon_A$ and $\epsilon_B$ all in meV with data from literature}
\begin{ruledtabular}
\begin{tabular}{ccccc}
& $a_{\|}\tilde{\beta_A}$& $a_{\|}\tilde{\beta_B}$ &$\epsilon_A$ & $\epsilon_B$\\
\hline
(this work)&$335(50)$&$1582(220)$&$186(30)$&$312(60)$\\
Exp.\cite{Ikonomov1}&$482(32)$&$1471(102)$&$144(3)$&$206(4)$\\
Exp.\cite{Giesen3}&$ $&$ $&$167$&$167$\\
Num.\cite{Feibelman1}&$ $&$ $&$180$&$250$\\
\end{tabular}
\end{ruledtabular}
\end{table}
\\
Thereon the step line stiffness $\tilde{\beta}$ has been calculated numerically for the $A$ and $B$ steps using Eq.~\ref{eq:ECS3}. The kink energies $\epsilon_{A,B}$ result from solving Eq.~\ref{eq:ECS5}.\\
The average values of the step line stiffness $\tilde{\beta}_{A,B}$ and the kink energies $\epsilon_{A,B}$ are given and compared to literature data in Tab.~\ref{tab:AFM-Paramters2}. The experimental step line stiffness of $A$ and $B$ steps obtained at $T=923$~K shows a significant anisotropy and is in the magnitude of stiffnesses obtained from vacancy island formed on Pt(111) single crystals at $T=713$~K~\cite{Ikonomov2}. Thereby $\tilde{\beta_A}$ is slightly smaller compared to the equivalent step line stiffness measured by Ikonomov \textit{et al.}~\cite{Ikonomov2} and $\tilde{\beta_B}$ slightly larger. This difference obviously refers to the enhanced anisotropy of the $A$ and $B$ step length in this work and to the different annealing temperature. It is assumed, that the found hillock shape represents the equilibrium shape in respect to the dominant configurational forces in this thin film geometry at $T=923$~K. Although the experimental framework in this work differs significantly from the one of Ikonomov \textit{et al.}~\cite{Ikonomov2}, a remarkable agreement of the step line stiffness is achieved.\\        
The same applies to the calculated kink energies $\epsilon_{A,B}$, which are in good agreement with  \textit{ab-initio} calculations~\cite{Feibelman1} as well as with the experimental results of Ikonomov~\cite{Ikonomov2}, see Tab.~\ref{tab:AFM-Paramters2}. In contrast to the data of Giesen \textit{et al.}~\cite{Giesen3}, the kink energy $\epsilon$ is significantly step-type dependent. The numerical values of the measured kink energies correspond well to the theoretical results of Feibelman~\cite{Feibelman1} at $T=0K$ and therefore cast the postulated temperature dependence of the kink energy~\cite{Kara1} into doubt.\\
However, although the absolute values for $\epsilon_{A,B}$ found in this work are slightly larger than in reference~\cite{Ikonomov2}, the ratios of the kink energies are with $\delta_\epsilon=\frac{\epsilon_{A}}{\epsilon_{B}}\approx0.5$ in good accordance. 
The numerical values of $\epsilon_{A,B}$ can now be used to determine the step depended bond energy $J_{A,B}=2\epsilon_{A,B}$ which serves as a key parameter in solid-on-solid KMC models~\cite{Dufay1} simulating thin film instabilities. It becomes obvious that if one wants to foresee and understand the fundamentals of thin film instabilities, it is inevitable to consider the general anisotropic character of the kink energy which directly affects the motion of atomic steps.
 
\section {\label{sec:level5}Conclusions}

In essence, it has been shown that the formation of hillocks in ion beam sputtered Pt thin films is a result of a complex stress relaxation process. The kinetics of the inherent mechanisms exhibit a different thermal activation and depend differently on the stress field inside the thin film. 
It is shown that two in general independent physical processes control the morphological evolution and kinetics of hillock formation: one is attributed the to minimization of the strain density $\Delta W$ and the other to the minimization of the surface energy $\Delta\Phi$. These two competing contributions to the total free energy cause a transition from pure hillock formation at $T\leq923$~K to a coexistence of hillock formation and film rupture in the vicinity of the hillock edges at $T\geq973$~K. The observed competition of relaxation mechanisms in this work is in excellent agreement with the predictions made by Tersoff~\cite{Tersoff1,Tersoff2}.\\
These findings are substantiated by the performed FEM simulations that clearly indicate that formation of hillocks on strained films cause a minimization of the thin films free energy as function of the hillocks aspect ratio $a$. The predicted aspect ratio that corresponds to the minimal free energy configuration of the thin film is in excellent agreement with experimentally measured mean aspect ratio. Therefore it is assumed that the observed hillocks possess equilibrium shape. In addition, it has been shown by FEM that the local maximum of the stress field in the thin film coincides with experimentally observed regions of rupture.\\  
The induced hillock formation is governed by the motion of atomic steps on the surface. The analysis of the hillock shapes by high resolution AFM revealed quantitatively two key parameters of the motion of atomic steps: the anisotropic step line stiffness $\tilde{\beta}_{A,B}$ and the anisotropic kink energy $\epsilon_{A,B}$. The found values are in good agreement with both experimental~\cite{Ikonomov2} and theoretical~\cite{Feibelman1} data from literature. On closer inspection, it can be seen that the hillock shape anisotropy found in this work slightly differs from the observed anisotropies in formed vacancy islands on Pt(111)~\cite{Ikonomov2} or grown Pt-islands~\cite{Michely2}. This is due to the different state of the surface and the presence of a substrate material in case of a thin film geometry.\\ 
It is noteworthy, that the choice of the deposition technique, ion beam sputtering in this work and magnetron sputtering in a previous work~\cite{Galinski1}, severely impacts the pathway of the thermal instability of the thin film. While in a previous study thermal annealing of the thin film led to hole formation at defect associated perturbations, the identical thermal treatment caused the formation of hillocks in the present work. Hence the inherent thin film properties like the distribution of grain boundary energies or the internal stress induced by the deposition technique have a decisive impact on the competing instability mechanisms and their subsequent shape changes.

\begin{acknowledgments}
The authors wish to acknowledge the financial supported by the Swiss Bundesamt f\"ur Energie (BfE) and Swiss Electric Research (SER) and would like to thank the EMEZ (Electron Microscopy Center, ETH Zurich) and Ulrich M\"uller (EMPA) for their support. In addition Henning Galinski thanks Iwan Schenker and Anna Evans for fruitful and stimulating discussions.
\end{acknowledgments}

\newpage 
\bibliographystyle{apsrev4-1}

\bibliography{Holes-Hillocks}

\end{document}